\documentclass[aps,pre,showpacs,preprint,groupedaddress,nofootinbib]{revtex4-1}
\usepackage{pdfsync}
\usepackage{ifpdf}
\ifpdf
\usepackage{hyperref}
\else
\usepackage[hypertex]{hyperref}
\fi
\usepackage{graphicx,subfigure}
\usepackage{amsmath,amsfonts,amssymb}
\usepackage{xcolor}

\newcommand{\f}{\frac}
\newcommand{\suml}{\sum\limits}

\newcommand{\intl}{\int\limits}

\graphicspath{{D:/localtexmf/Mytex/LGH/mathplots/}}

\begin{document}
\title{The unified picture for the
classical laws of Batschinski and the rectilinear diameter for
molecular fluids}
\author{L.~A. Bulavin}
\email{bulavin@univ.kiev.ua}
\affiliation{Department of Molecular Physics, \\
Taras Shevchenko National University of Kyiv, 2, Prosp.
Academician Glushkov, Kyiv 03022, Ukraine}
\author{V.~L.~Kulinskii}
\email{kulinskij@onu.edu.ua}
\affiliation{Department of Theoretical Physics, Odessa National
University, Dvoryanskaya 2, 65026 Odessa, Ukraine}
\begin{abstract}
The explicit relations between the thermodynamic functions of
the Lattice Gas model and the fluid within the framework of
approach proposed earlier in [V.~L. Kulinskii, J. Phys. Chem. B
\textbf{114} 2852 (2010)] are derived. It is shown that the
Widom line serves as the natural border between the gas-like
and the liquid-like states of the fluid. The explanation of the
global cubic form of the binodal for the molecular liquids is
proposed and the estimate for the amplitude of the binodal
opening is obtained.
\end{abstract}
%
\maketitle
\section{Introduction}\label{sec_intro}
There are two classical laws for the molecular liquids known
more than a century. Long time they were considered as mere
curious facts restricted to the simple van der Waals equation
of state (vdW EoS). First of them is the law of the rectilinear
diameter (LRD) \cite{crit_diam0,*crit_diam1_young_philmag1900,
*crit_diambenzene_physrev1900,pcs_guggenheim_jcp1945}. It
states that the diameter of the coexistence curve in terms of
density-temperature is the straight line:
\begin{equation}\label{diameter_class}
  \tilde{n}_{d} = \f{n_l+n_g}{2n_c}  = 1+
 A\,\f{T_c-T}{T_c}\,\,\,, \quad A>0
\end{equation}
where $n_{i},\,i = l,g$ are the densities of the liquid and the
gas phases correspondingly, $n_c$ is the critical density, $T$
is the temperature. Further we put the Boltzmann constant to
unit: $k_B = 1$. Another simple linear relation is the
Batschinski law \cite{eos_zenobatschinski_annphys1906} for the
vdW EoS:
\begin{equation}\label{p_vdw}
  P = \f{n\,T}{1-n\,b} - a\,n^2\,,
\end{equation}
where $P$ - is the pressure. It states that the line determined
by the condition $Z=1\,,$ where $Z = P/(n\,T)$ is the
compressibility factor is the straight line:
\begin{equation}\label{batschinski_vdw}
  \f{n}{n_{*}}+\f{T}{T_{*}} = 1\,.
\end{equation}
Here $n_{*} = 1/b$, $T_{*} = a/b$ is the Boyle temperature in
the van der Waals approximation and $a,b$ are the parameters of
the vdW EoS (see e.g.~\cite{book_hansenmcdonald}). In general
case of the spherically symmetrical potential $T_*$ is
determined in accordance with \cite{book_ll5_en} as following:
\begin{equation}\label{tbvdw}
  T^{(vdW)}_B  = \f{a}{b}\,.
\end{equation}
where
\begin{align}\label{vdw_ab}
a =\,\, -2\pi\,\intl_{\sigma}^{+\infty}\Phi_{attr}(r)\,r^2\,dr
\end{align}
and $\Phi_{attr}(r)$ is the attractive part of the full
potential $\Phi(r)$, $\sigma$ is the effective diameter of the
particle so that $b = \f{2\pi}{3}\,\sigma^{3}$. The definition
for the density parameter $n_*$ will be given below.

The connection between two linear relations
\ref{diameter_class} and \eqref{batschinski_vdw} has attracted
attention after work of Herschbach\&Coll.
\cite{eos_zenobenamotz_isrchemphysj1990,eos_zeno_jphyschem1992},
where the line $Z=1$ was named by the Zeno-line and it was
shown that it is indeed almost straight for the normal fluids
though the deviations are noticeable for the accurate data. The
straightness of the Zeno-line implies the constraint on the
contact value of the radial distribution function
\cite{eos_zenoline_jphyschem1992}. Moreover, there are
correlations between these linear elements and the locus of the
critical point which were discovered in series of works of
Apfelbaum and Vorob'ev
\cite{eos_zenoapfelbaum_jpchema2004,eos_zenoapfelbaum_jpchemb2006,
eos_zenoline_potentials_jcp2009}. The authors put forward the
creative idea about the tangency of the Zeno-line to the
liquid-vapor binodal extrapolated to the nonphysical region
$T\to 0$. This allowed to connect the locus of the CP with the
parameters of the Zeno-line:
\begin{equation}\label{zeno_apfvor}
  \f{T}{T_B}  + \f{n}{n_B} = 1\,.
\end{equation}
which are $T_B$ and $n_B$. They determine the intersection
points for \eqref{zeno_apfvor} with the corresponding axes.
According to former treatments
\cite{eos_zenobenamotz_isrchemphysj1990,eos_zeno_jphyschem1992,
eos_zenoapfelbaum_jpchema2004,eos_zenoapfelbaum_jpchemb2006},\,
$T_B$\, is the Boyle temperature, i.e.
\[B_{2}(T_{B})= 0\,,\]
and $n_B$ is determined by the relation:
\begin{equation}\label{nb}
n_B= \f{ T_B }{B_3\left(\,T_B\,\right)}\,\left. \f{dB_2}{dT}\right|_{T= T_B}\,.
\end{equation}
where $B_n$ is the virial coefficient of $n$-th order
\cite{book_hansenmcdonald}.

The relation \eqref{diameter_class} fails at the vicinity of
the critical point (CP) where the singular terms appear
\cite{book_patpokr,crit_rehrmermin_pra1973} (see also recent
works
\cite{crit_fishmixdiam1_pre2003,crit_aniswangasymmetry_pre2007,
crit_can_diamsing_kulimalo_physa2009}). Nonlinear deviations
from the linearities \eqref{diameter_class} and
\ref{zeno_apfvor} in low temperature region, e.g. for water can
be associated with the presence of the anisotropic interactions
like H-bonds \cite{water_zenoline_ijthermophys2001}. Despite
the deviations from the exactly linear behavior the relations
\ref{diameter_class} and \eqref{batschinski_vdw} are much more
general and can be considered as the nontrivial extension of
the principle of corresponding states
\cite{eos_zeno_jphyschem1992}.

In \cite{eos_zenome_jphyschemb2010} it was shown that both
\ref{diameter_class} and \eqref{batschinski_vdw} as well as the
correlations between these linear elements and the locus of the
critical point can be considered as the consequence of the
global mapping between the liquid-vapor part of the phase
diagram and that of the Lattice Gas. The latter is described by
the Hamiltonian:
\begin{equation}\label{ham_latticegas}
  H = -J\suml_{
\left\langle\, ij \,\right\rangle
  } \, q_{i}\,q_{j} - h\,\suml_{i}\,q_{i}\,.
\end{equation}
Here $q_{i}$ is the site filling number and $q_{i} = 0,1$
whether the site is empty or occupied correspondingly. The
quantity $J$ is the energy of the site-site interaction of the
nearest sites $i$ and $j$, $h$ is the field conjugated to the
filling variable $q_i$. We denote by $t$ the temperature
variable corresponding to the Hamiltonian
\eqref{ham_latticegas}. The density parameter is the
probability of occupation of the lattice site $x =
\left\langle\, q_i \,\right\rangle$.

If the LRD is assumed then such mapping is given by:
\begin{equation}\label{projtransfr_nx}
  n =\, n_*\,\f{x}{1+z \,t}\,,\quad
  T =\, T_*\,\f{z\, t}{1+z \,t}\,,
\end{equation}
where $z$, $n_*$ and $T_*$ are some parameters, which are
connected with the coordinates of the CP:
\begin{align}\label{cp_fluid}
      z =& \, \f{T_c}{T_* - T_c}\,,\\
  n_{c} =& \, \f{n_*}{2\left(\,1+z\,\right)}\,,\\
  T_{c} =& \, T_*\, \f{z}{1+z}\,.
\end{align}
This transformation is uniquely determined by the
correspondence between the characteristic \emph{linear}
elements on the phase diagrams of the fluid and the LG. It is
assumed that the coordinates of the CP for the LG are
normalized so that $t_c = 1$ and $x_c = 1/2$. In such a context
the parameter $z$ represents the class of the corresponding
states \cite{eos_zenogenpcs_jcp2010}. Note that at $z\to 0$ and
$T_{*}\to \infty $ with $z\,T_{*}\to 1$ from
\ref{projtransfr_nx} we get $n/n_{*} \to x$ and $T \to t$.

The application of the transformation to the calculation of the
locus of the critical point of Lennard-Jones fluids is given in
\cite{eos_zenogenpcs_jcp2010,eos_zenomeglobal_jcp2010}. The
strong argument in favor of the choice of the linear element
\ref{batschinski_vdw} with the parameters $n_*, T_*$ instead of
those for the Zeno-line \eqref{zeno_apfvor} and the comparison
of the values for various potentials are given in
\cite{eos_zenogenpcs_jcp2010,crit_globalisome_jcp2010}. There
\ref{projtransfr_nx} was used to map the binodal of the planar
Ising model onto the binodal of the two-dimensional
Lennard-Jones fluid.

The aim of this paper is to discuss physical basis of the
linearities \eqref{diameter_class} and \eqref{zeno_apfvor} on
the liquid-vapor part of the phase diagram of the fluids. We
will follow the results of
\cite{eos_zenome_jphyschemb2010,eos_zenogenpcs_jcp2010}. We
expand some arguments of \cite{eos_zenogenpcs_jcp2010}, which
concern the corrected interpretation of the Batchinski law in a
way consistent with the van der Waals approximation for the
EoS. Also we derive the relation between the thermodynamic
potentials for the continuum and the lattice models of fluids.
\section{Liquid-vapor binodal as the image of the binodal
of the lattice model} It is easy to see that linear laws
\ref{batschinski_vdw} and \eqref{zeno_apfvor} are fulfilled
trivially in the case of the Lattice Gas (LG) model or
equivalently the Ising model. Indeed, the rectilinear diameter
law for the LG is fulfilled due to the symmetry of the
Hamiltonian \eqref{ham_latticegas} with respect to the line $x
= 1/2$.

The analog of the Zeno-line for the LG can be defined too. In
this case it is the line $x=1$ where the ``holes`` are absent.
This is consistent with the basic expression for the
compressibility factor \cite{book_hansenmcdonald}:
\begin{equation}\label{zpt}
  Z =\f{P}{n\,T} = 1-\f{2\pi \,n}{3\,T}\int r^3
\frac{\partial\, \Phi(r)}{\partial\, r}\,g_2(r\,;n,T)\, d\,r\,,
\end{equation}
and the definition of the Zeno-line as the one where the
correlations determined by the repulsive and the attractive
parts of the potential compensate each other. It is clear that
if $x=1$ then the perfect configurational order takes place and
the site-site correlation function of \eqref{ham_latticegas}
vanishes:
\[
\left\langle\,
\left\langle\, q_i\,q_j \,\right\rangle
 \,\right\rangle
 =
\left\langle\, q_{i}\,q_{j} \,\right\rangle
  -  \left\langle\, q_{i}\,\right\rangle\left\langle\, q_{j} \,\right\rangle=0\,\,.\]
Thus the line $x=1$ plays the role of the Zeno-line on the
$x-t$ phase diagram of the LG. Due to simple structure of the
LG Hamiltonian \eqref{ham_latticegas} and explicit symmetries
there are degenerate\label{degeneration} elements of the phase
diagram. The critical isochore  $x_{c} = 1/2$ coincides with
the diameter. The Zeno-line is the tangent to the binodal which
in this case expands up into the region $t\to 0$. Thus there is
the degeneration of these in the case of the LG. Naturally,
that both mentioned degenerations for the linear elements
\ref{diameter_class} and \eqref{batschinski_vdw} of the phase
diagram disappear for the real fluids. The difference of the
diameter and the isochore is nothing but the asymmetry of the
binodal \cite{crit_2betasengers_engchem1970}. The difference
between the Zeno-line and the tangent to the extrapolation of
the binodal into the low temperature region $T\to 0$ takes
place only for the vdW EoS and influences directly the approach
of
\cite{eos_zenoapfelbaum_jpchemb2006,eos_zenoapfelbaum_jpchemb2008}.
The latter is based heavily of the constraint of the tangency
to the extrapolation of the binodal. In fact there is no
physical reasons to identify such tangent line with the
Zeno-line. This is possible only for the vdW EoS. Using the
generalized van der Waals approach of
\cite{eos_genvdw_jcp2001,*eos_genvdw_jpchem2003} any EoS can be
approximated by the vdW EoS with the corresponding parameters.
The definition of the tangent linear element
\ref{batschinski_vdw} relies upon the van der Waals
approximation for the given EoS and therefore does not coincide
with the Zeno-line. The value of $n_*$ is determined by the
condition analogous to \eqref{nb}:
\begin{equation}\label{nbme}
n_*= \f{ T_* }{B_3\left(\,T_*\,\right)}\,\left. \f{dB_2}{dT}\right|_{T= T_*}\,.
\end{equation}
This relation follows from the constraint:
\begin{equation}\label{zme}
\f{d}{d\,T}\,\left(\,\f{Z(n(T),T) - 1}{n(T)} \,\right) = 0
\end{equation}
which generalizes the condition $Z = 1$ and implies the linear
change of the compressibility factor with the temperature $T$
along the linear element \eqref{batschinski_vdw}.

The proposed simple relation \eqref{projtransfr_nx} between the
LG and the fluid may be useful for construction of the
empirical EoS for the real substances. Recently, the nonlinear
generalization of \eqref{projtransfr_nx} has been proposed in
\cite{eos_zeno_lattice2real_jpcb2010}:
\begin{equation}\label{av_nonlin}
  n =\, n_*\,\f{x^{\gamma}}{1+z \,t}\,,\quad
  T =\, T_*\,\f{z\, t}{1+z \,t}\,,
\end{equation}
with $\gamma$ as the fitting parameter. Unfortunately, the
physical meaning of the parameter $\gamma$ and its connection
with the interaction potential was not discussed. Moreover from
the basic thermodynamical reasonings the extensive parameters
such as number of particles in the LG  and in fluid should be
proportional unless the fluctuational effects are taken into
account. The lasts are the source of the fluctuational induced
shift of the mean-field position of the critical point
\cite{book_ma}. So the modification of the
\eqref{projtransfr_nx} in order to obtain the exact position of
the critical point should be based on the inclusion of the
fluctuation effects and their scaling properties. In
\cite{eos_zenogenpcs_jcp2010} it was shown how the relations
\ref{projtransfr_nx} augmented with some scaling considerations
allow to obtain the critical points of the Lennard-Jones fluids
basing on the properties of the potentials. The difference
between $T_{B}$ and $T_*$, $n_B$ and $n_*$ is indeed essential
especially in high dimensions $d>3$
\cite{eos_zenomeglobal_jcp2010}.

In the next Section we propose the generalization of the
transformation \eqref{projtransfr_nx} for the procedure of the
symmetrization of the binodal of the fluid which preserve the
correspondence between the extensive thermodynamic quantities.

\section{Symmetrization of the binodal}\label{sec_symbinodal}%
We believe that the key point which determines the form of the
mapping is the LRD \eqref{diameter_class}. The linear element
\ref{batschinski_vdw} plays auxiliary role and defines the
proper scales for the density and the temperature.

As was shown earlier in \cite{crit_globalisome_jcp2010} the
transformation \eqref{projtransfr_nx} allows to map the binodal
$T_{bin}(n)$ of the Lennard-Jones fluid onto the binodal
$t_{bin}(x)$ of the corresponding lattice gas model. The latter
has explicitly symmetric shape with respect to
the critical isochore $x_c=1/2$: %
\begin{equation}\label{tbx_even}
t_{bin}(\tilde{x}) = t_{bin}(-\tilde{x})\,,\quad \tilde{x} = 1/2-x\,,\,\,
x\le 1/2
\end{equation}
due to the particle-hole symmetry of the Hamiltonian
\ref{ham_latticegas}. Thus one can treat \eqref{projtransfr_nx}
as the procedure of the symmetrization of the phase diagram.
Indeed, suppose that $n_d(T)$ is the dependence of the density
diameter. Then the variable $n/n_d(T)$ is symmetrical over the
binodal. With this \eqref{projtransfr_nx} may be generalized as
following:
\begin{equation}\label{projtransfr_nx_gen}
  x = \f{n}{2\,n_{d}(T)}\,,\quad
T/T_*\, = 1-f(t)\,,
\end{equation}
where the parametrization function $f$ is chosen so that to map
the binodal of the fluid $T_{bin}(n)$ onto the binodal of the
LG $t_{bin}(x)$. It can be found from the common  conditions of
the thermodynamic equilibrium: $P(n_{g},T) = P(n_{l},T)$. The
equality of the chemical potentials is fulfilled due to
symmetry of the binodal of the LG (see also
Section~\ref{sec_trmdfunctions}).


The transformation \eqref{projtransfr_nx} is the particular
case for which:
\[n_{d}(T) = \f{n_*}{2} f(t)\,,\quad f(t) = \f{1}{1+z\,t}\]

\begin{figure}[hbt!]
\center
  \includegraphics[scale=0.85]{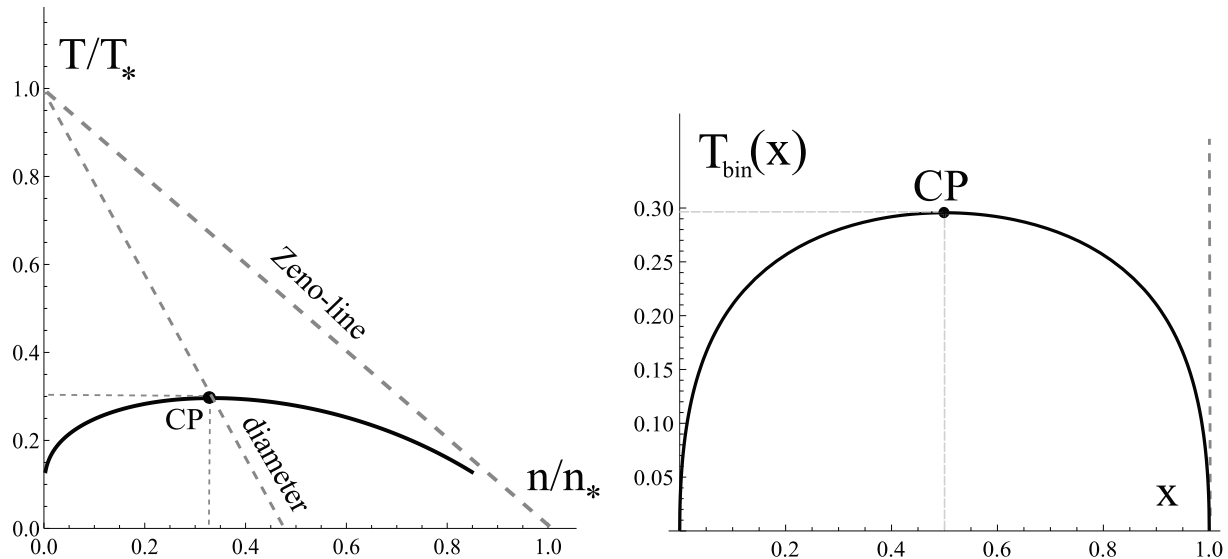}\\
  \caption{The symmetrization of the vdW binodal using the parametrization \eqref{linear_parametriz} corresponding to the linear approximation for the diameter.}\label{fig_ising_vdw}
\end{figure}

In the simplest \emph{linear} approximation for the temperature
behavior of the diameter $n_d$ of the form
\ref{diameter_class}, obviously:
\begin{equation}\label{linear_parametriz}
n/n_* = x\,f(t)\,,\quad T/T_*\, = 1-f(t)\,\,.
\end{equation}
The unknown parametrization function $f$ is determined by the
condition
\begin{equation}\label{plpg}
P\left(n_l,T\right) = P\left(n_g,T\right)\,.
\end{equation}
To illustrate this procedure let us consider the classical vdW
EoS \eqref{p_vdw}. Using the symmetry of the LG representation
\ref{linear_parametriz} with respect to $x_c = 1/2$ it is
convenient to represent the densities of the coexisting phases
as $n_{g} = x\,f(t)$ and $n_{l} = (1-x)\,f(t)$, $0\le x\le
1/2$. Substituting these relations into \eqref{plpg} where the
pressure $P$ is given by \eqref{p_vdw} we get simple algebraic
equation for the value $f$:
\begin{equation}\label{f_eq}
  1 - 2 f + f^2 + f^3\,(x-1)\,x = 0\,,
\end{equation}
as a function of $x$. In accordance with
\eqref{linear_parametriz} this provides the symmetrization of
the binodal of the vdW EoS in terms of the LG variable $x$ The
result is shown in Fig.~\ref{fig_ising_vdw}. The corresponding
coordinates of the critical point in accordance with
\ref{linear_parametriz} are:
\begin{equation}\label{vdw_cp_linear}
  n_c/n_* = \f{1}{2}\,f(1) = 0.352\,,\quad T_c/T_* = 1-f(1)\approx 0.296\,,\quad P_c/(n_*\,T_*)\approx 0.037\,.
\end{equation}
The difference between the exact values $n_c/n_*  = 1/3,
T_{c}/T_{*} =8/27, P_c/(n_*T_*) = 1/27$ is caused by the
deviation of the diameter for the vdW EoS form the linear
behavior. As we see the differences are rather\, small. Thus
simple parametrization \eqref{linear_parametriz} can be applied
to any EoS where the deviation from the LRD can be neglected.
In this case the symmetrization of the binodal described by the
parametrization \eqref{projtransfr_nx_gen} or
\ref{linear_parametriz} along with the EoS for the LG could be
useful in processing the data of the simulations for the
coexistence curve (CC) of the Lennard-Jones fluids, where
simple approximate formula:
\begin{equation}\label{cubicextrapol}
n_{l,g} = n_c+A\,|\tau|\pm B_0\,|\tau|^{\beta}\,,\quad \tau  = \f{T-T_c}{T_c}\,,
\end{equation}
is used \cite{book_frenkelsimul}. Also the proposed approach
allows to avoid ambiguity in extrapolation of the binodal into
the region $T\to 0$ \cite{eos_zenoapfelbaum_jpchemb2006}.

\begin{figure}
\center
 \subfigure[]{\includegraphics[scale=0.35]{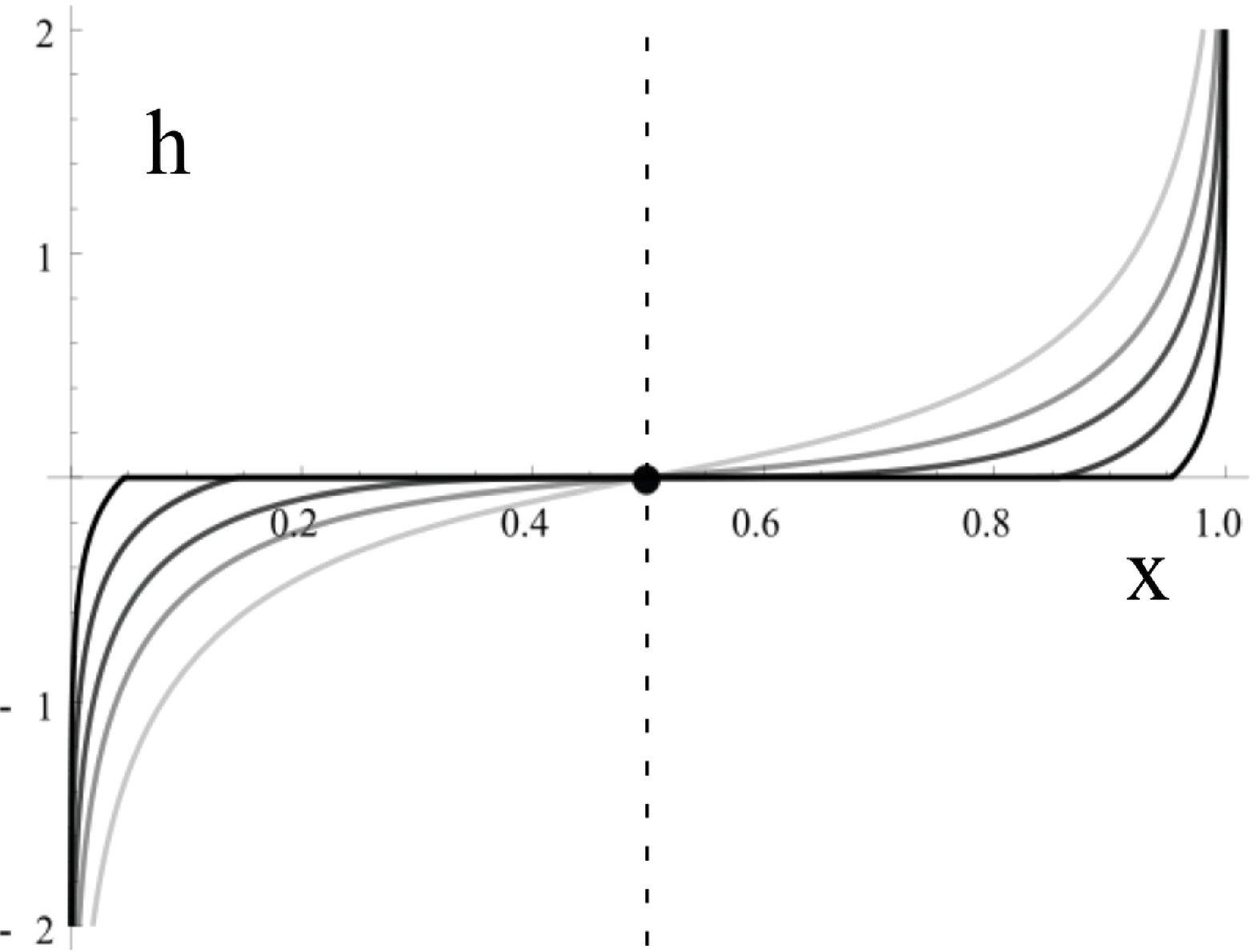}}\,\,
 \subfigure[]{\includegraphics[scale=0.65]{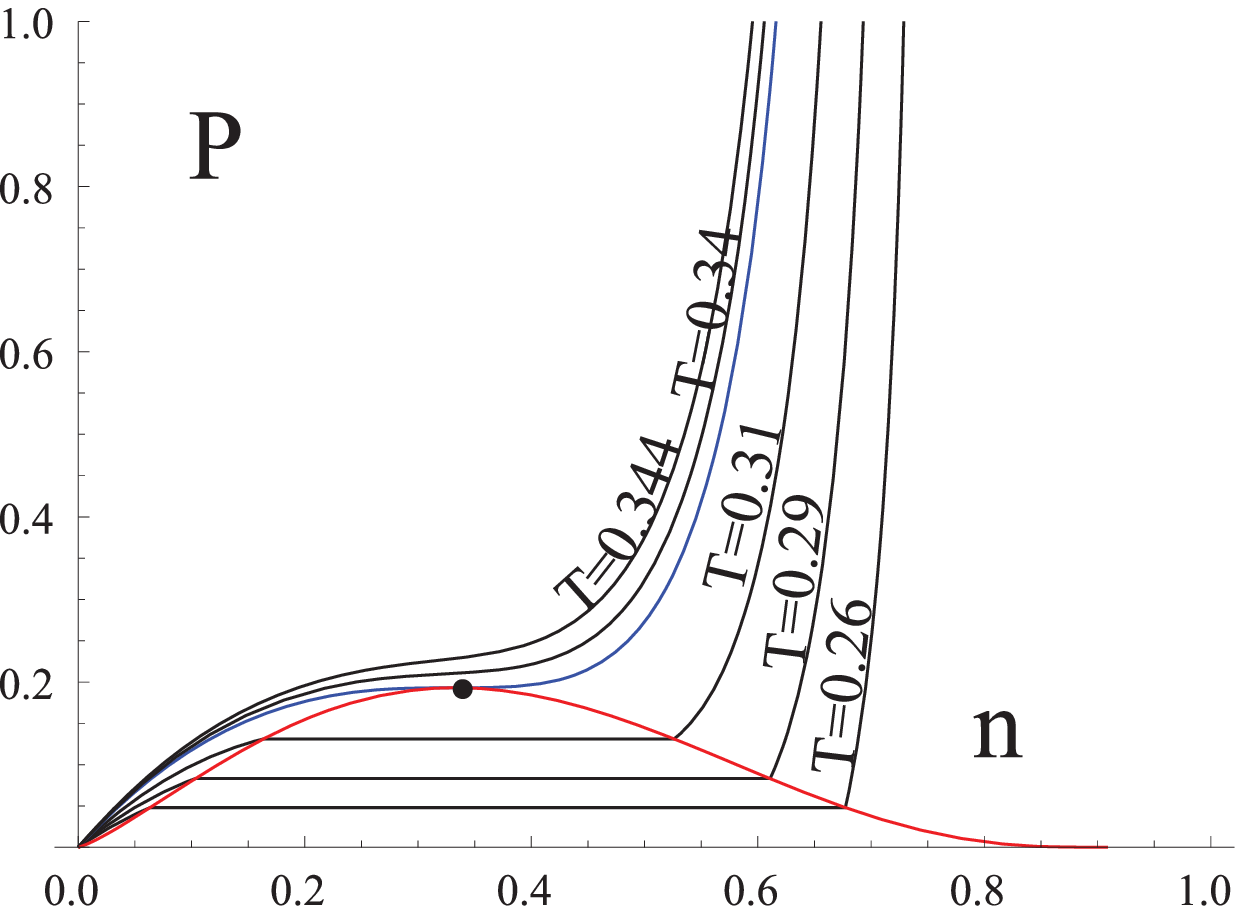}}\\
  \caption{The isotherms obtained using \eqref{projtransfr_nx} applied to Curie-Weiss approximation for LG with $z=1/2$. The isotherms of \eqref{cw_eos} are shown in Fig.~(a). The corresponding isotherms for the isomorphic ``fluid`` model are in Fig.~(b).
The red curve is the binodal, the critical isotherm $T_c = 1/3$ is marked by the blue line. The quantities are dimensionless (see \eqref{vdw_cp_linear}).}\label{fig_p_widomline}
\end{figure}

In order to visualize how the splitting of the degenerate
elements mentioned above (see p.~\pageref{degeneration}) occurs
it is expedient to consider the application of the
transformation to the classical EoS for the LG the Curie-Weiss
molecular field approximation (see e.g.
\cite{book_huang_statmech}). This equation of state has the
form\footnote{Here we neglect the trivial difference between
the $m$-field conjugated to variable $m$ and the $x$-field
conjugated to $x$ because it is irrelevant for our
consideration.}:
\begin{equation}\label{cw_eos}
h(m,t) = t\,{\rm ArcTanh}\,(m) - m \,,\quad m = 2x-1\,.
\end{equation}
The isotherms for this model EoS are shown in
\ref{fig_p_widomline}~(a). The linear character of the
transformation \eqref{projtransfr_nx} with respect to the order
parameters $x$ and $n$ allows to connect the pressure for the
asymmetrical liquid $P_{liq}(T,\mu)$ with that of the LG as
follows:
\begin{equation}\label{pp}
  P_{liq}(T,\mu) = P_{LG}(t(T),h(\mu,T))\,.
\end{equation}
The dependence $h(\mu,T)$ will be obtained in
Section~\ref{sec_trmdfunctions} (see \eqref{muh} below). The
pressure for the LG $P_{LG}$ and in particular for EoS
\ref{cw_eos} can be defined by the standard method (see e.g.
\cite{book_rice_thermodyn}). Taking into account the
conjugation of the thermodynamic variables we can write:
\begin{equation}\label{p_lg}
P_{LG}(t,h) = f(t,m)+\f{m+1}{2}\,h\,,
\end{equation}
where $f(t,m)$ is the thermodynamic potential for the variables
$(t,m)$. It is constructed easily using EoS \eqref{cw_eos} so
that:
\begin{equation}\label{f_mt}
  h(t,m) = \left.
  \frac{\partial\, f}{\partial\, m}\right|_{t}\,.
\end{equation}

Substituting \eqref{p_lg} into \eqref{pp} we are able to
construct the isotherms in $n-T$ plane. The result of
construction of the isotherms and the binodal of the ``fluid``
in coordinates ``pressure-density`` basing on the Curie-Weiss
EoS \eqref{cw_eos} for the LG is shown in
\ref{fig_p_widomline}.

In the following Section we discuss the relation between the
thermodynamic functions of the LG and its continuum analog in
detail.
\section{The relation between the thermodynamic potentials of
the fluid and the Lattice Gas} \label{sec_trmdfunctions}%
Basing on the analysis of
\cite{eos_zeno_lattice2real_jpcb2010,crit_globalisome_jcp2010}
one can expect that the transformation \eqref{projtransfr_nx}
gives the relation between phase diagram of the LG and that of
the real fluid at least as the ``zeroth order approximation``.
The idea that the difference between irregularity configuration
for continuum fluids and regularity of configurations of
lattice models is unimportant for consideration of the
order-disorder transitions in the fluctuational region is due
to K.S. Pitzer (see \cite{eos_pitzer_purapplchem1989}). But
beyond the fluctuational region the shape of the holes in real
or continuum liquid and that in the lattice gas causes the main
difference between the configurations of these systems. From
this point of view the global character of the transformation
\ref{projtransfr_nx} shows that the particle-hole simplified
picture still can be useful. Though, it is not the particle
density which reflects such symmetry. Rather the combination of
the density of the particles and the density of the holes is
the symmetrical variable (see p.~\pageref{page_symmorder}).
This rehabilitates the hole theory for expanded liquids
\cite{book_rice_thermodyn,liq_barkerhenderson_rmp1973}. Such
caricature picture of the liquid state gives the possibility to
relate the thermodynamic functions of these systems.

Let
\[\mathfrak{G}(t,h,\mathcal{N}) = \mathcal{N}\,\mathfrak{g}(t,h)\,,\quad
\text{and}\,\quad J(T,\mu,V) = P(T,\mu)\,V\,,\] are the
thermodynamic potentials of the grand canonical ensembles for
the LG and the fluid correspondingly. Here $\mathcal{N}$ is the
number of sites in a lattice. First it is natural to state the
following relation $\mathcal{N} = n_* \,V$ between the
extensive variables of these ensembles. The results of
~\cite{eos_zenoapfelbaum_jpchema2008} allow to interpret
$1/n_*$ as the volume per particle in the ideal crystal state
at $T\to 0$. Such a state defines the lattice which may serve
as the basis for the determination of the corresponding Lattice
Gas model.

Using the standard definitions:
\begin{equation}\label{nx_definitions}
n =\f{1}{V}\left.\frac{\partial\, J}{\partial\, \mu}\right|_{T}\,,\quad x
=\f{1}{\mathcal{N}} \left.\frac{\partial\, \mathfrak{G}}{ \partial\,
h}\right|_{t}
\end{equation}
along with \eqref{projtransfr_nx} we get the following relation
between the potentials:
\begin{equation}\label{transfm_potentials}
J (\mu,T,V) =  \mathfrak{G}
\left(\,h(\mu,T),t(T), \mathcal{N}
\,\right)\,\Rightarrow P(\mu,T) = n_{*}\,\mathfrak{g}
\left(\,h(\mu,T),t(T)\,\right)\,.
\end{equation}
From \eqref{projtransfr_nx} the relation between the density of
the fluid and the density of the LG can be written as
following:
\begin{equation}\label{nmut}
  n(\mu,T)/n_* = x(h(\mu,T),t(T))\,\left(\,1-T/T_* \,\right)\,.
\end{equation}
Taking into account trivial relation:
\[\left.\frac{\partial\, }{\partial\, \mu}\right|_{T} =
\left.\frac{\partial\, h }{\partial\, \mu}\right|_{T}
\left.\frac{\partial\, }{\partial\, h}\right|_{t}\,,\] from
\eqref{nx_definitions},\ref{transfm_potentials} and
\eqref{nmut} we get the following relation:
\begin{equation}\label{hmu}
  h(\mu,T) = \left(\,1-T/T_* \,\right)
  \left(\,\mu -\mu_{0}(T) \,\right)\,,
\end{equation}
or, in the inverse form:
\begin{equation}\label{muh}
  \mu -\mu_0(T)  = h\,\left(\,1+z\,t \,\right)\,\,.
\end{equation}
We remind that $h=0$ below CP is the coexistence line for the
LG and is mapped onto the saturation curve of the continuum
fluid. Therefore $\mu_0(T)$ coincides with the chemical
potential $\mu_{s}(T)$  along the saturation curve below the
critical point $T<T_c$. To determine $\mu_0(T)$ in the
supercritical region $T>T_{c}$ we note that the line
\begin{equation}\label{musupercrit}
  n(\mu_{0}(T),T)/n_* = \f{1-T/T_*}{2}\,,
\end{equation}
is the image of the line of symmetry $h=0$ for the LG along
which $x=1/2$.
\begin{figure}[hbt!]
\center
  \includegraphics[scale=1]{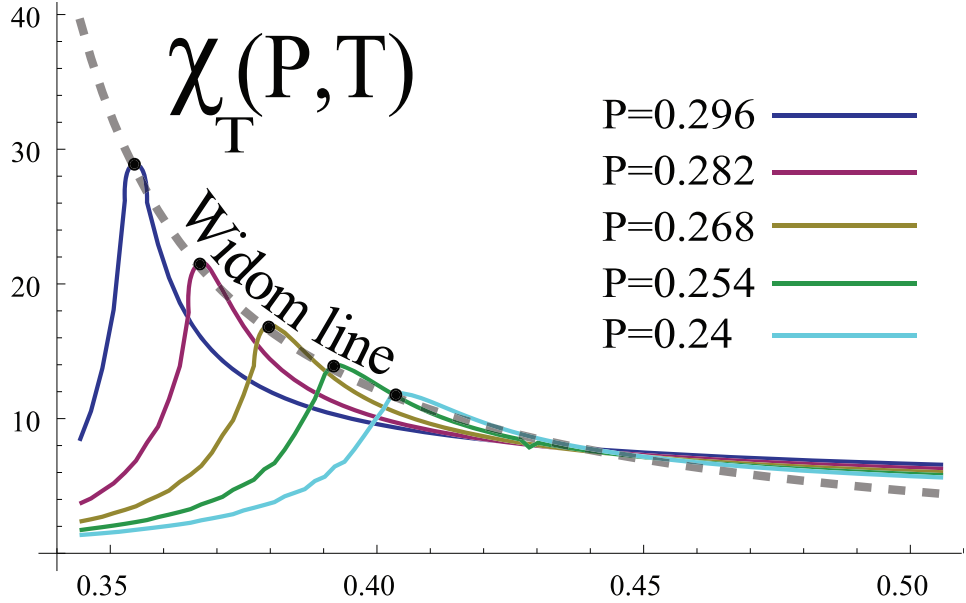}\\
  \caption{Widom line (dashed) as the line of the temperature maxima of the isothermal  compressibility $\chi_{T}(P,T)$ of the fluid for different $P$. It is given by \eqref{musupercrit}. The compressibility is calculated using the Curie-Weiss EoS \eqref{cw_eos} with subsequent transformation \eqref{projtransfr_nx}.}\label{fig_compress_widomline}
\end{figure}
Therefore $\mu_0(T)$ defined by \eqref{musupercrit} can be
considered as the Widom-Stillinger line of symmetry for the
liquid
\cite{crit_diamrowidom_jcp1970,*crit_diampartholewidom_jcp1973}.
Indeed, such line is defined as the locus of maximum of
correlation length where the thermodynamic response function
such as the isothermal compressibility $\chi_{T} =
\f{1}{n}\,\left. \frac{\partial\, n}{\partial\, p}\right|_{T}$
has maxima $T_{max}(P)$ \cite{fwidomline_jpcm2007}. Obviously,
in case of the LG it is represented by the line $h=0$, or
equivalently by the critical isochore $x=1/2$.

The Widom-Stillinger line has attracted much attention recently
in the studies on supercritical states of liquids
\cite{crit_supercrit_prl2006,crit_widomline_nature2010}. The
stated mapping between the LG and the fluid naturally explains
the separation of the fluid region into the gas-like and the
liquid-like states with the Widom line as the border. The
gas-like region is the image of the region $x<1/2$ where the
number of holes is greater than the number of particles. For
the liquid-like region $x>1/2$ the situation is inverse.

The line \eqref{musupercrit} is the image of the line $h=0$.
Therefore along this line the continuation of the subcritical
behavior (divergence) of the isothermal  compressibility
$\chi_{T}$ takes place. Fig.~\ref{fig_compress_widomline} shows
the result of the calculations for the compressibility
$\chi_{T}$ in accordance with the relation
\ref{transfm_potentials} between the fluid and the LG. This is
the locus of the compressibility maxima.

From \eqref{transfm_potentials} the relation between the
entropy $\mathfrak{S}$ of the LG and that of the fluid $S$ can
be derived:
\begin{equation}\label{transfrm_entropy}
  S(T,\mu,V)  =
  \left.\frac{\partial\, J}{\partial\, T}\right|_{\mu} =
  \f{d\, t}{d\,T}
\left.\frac{\partial\, \mathfrak{G}}{\partial\, t}\right|_{h} =
\f{\left(\,1+z\,t\,\right)^2}{z}\,\mathfrak{S}(t,h,\mathcal{N})\,.
\end{equation}

The results described above illustrate the general analysis of
\cite{crit_rehrmermin_pra1973}. The mapping
\ref{projtransfr_nx} determines the fluid as the asymmetric
model for the symmetric LG. The conservation of the RDL is due
to the direct relation between the the thermodynamic potentials
\ref{transfm_potentials}. Indeed, if linear relation between
$P(\mu,T)$ and $\mathfrak{g}(h(\mu ,T),t(\mu,T))$ with the
analytic coefficients holds, then the quantity
$\left.\frac{\partial\, P(\mu,T)}{\partial\, h}\right|_{T}$ has
the rectilinear diameter. This order parameter is the
combination of the density and the entropy
\cite{crit_rehrmermin_pra1973,crit_aniswangasymmetry_pre2007}.

In order to be physically meaningful, the transformation
\ref{projtransfr_nx} or its generalization should appear as the
average of the transformation of the microscopic variables. We
assume that there exist such transformation which leads to the
symmetrization of the binodal. The similarity between the
liquid-vapor part of the diagram for the fluid and that of the
LG and also the separation between the liquid-like and gas
-like states (see e.g. \cite{crit_widomline_nature2010}) allows
to state that for the case of fluids there exists the
``asymmetrical`` microscopic observable $\mathcal{A}$ such
that:\label{page_symmorder}
\begin{equation}\label{adiam}
\left\langle\, \mathcal{A} \,\right\rangle_{l}+\left\langle\, \mathcal{A}
\,\right\rangle_{g} = 0\,.
\end{equation}
Here $ \left\langle\, \ldots \,\right\rangle_{l,g} $ stands for
the averages on the corresponding coexisting liquid and gaseous
states with $(T,n_{l}(T))$ and $(T,n_{g}(T))$.  In terms of
\cite{crit_griffitswheeler_pra1970} the equilibrium average
$\left\langle\, \mathcal{A} \,\right\rangle_{l,g} \ne 0$ is the
density-like variable, which takes different values in the
coexisting phases. The specific form of the observable depends
on the Hamiltonian of the fluid. Such observable exists in case
of penetrable sphere models where the relation between the
thermodynamic potentials $P$ and $\mathfrak{g}$ is linear
\cite{crit_rehrmermin_pra1973}. In fact the right hand side in
\ref{adiam} can be any analytic function of $T$ (including the
neighborhood of the CP). In such case it is possible to
redefine $\mathcal{A}$ so that \eqref{adiam} is fulfilled. Then
the following equation
\begin{equation}\label{a_0}
  \left\langle\, \mathcal{A} \,\right\rangle_{\mu,T} = 0\,\,,
\end{equation}
determines the continuation $\mu_0(T)$ of the diameter into the
supercritical region $T>T_c$. Once such observable is
determined the corresponding thermodynamic potential can be
defined so that
\[\left\langle\, \mathcal{A} \,\right\rangle= \left.
\frac{\partial\, \mathfrak{F}}
{\partial\, h_{\mathcal{A}}}\right|_{S}\,\,.\]

In general $\mathfrak{F}$ depends on the field variables
$P,\mu,T$ nonlinearly. Therefore, in accordance with
\cite{crit_griffitswheeler_pra1970,crit_rehrmermin_pra1973} the
density diameter shows both $1-\alpha $ and $2\beta$ anomalies.
Since the singularity of the diameter is of fluctuational
nature the transformation \eqref{projtransfr_nx} should be
considered as the mean-field approximation for the
transformation between microscopic fields
(see~\cite{crit_can_diamsing_kulimalo_physa2009}).

\section{The cubic shape of the binodal}
The fact of the global cubic shape of the CC for the molecular
liquids is the long standing issue and was well known to van
der Waals due to studies of Verschaffelt
\cite{crit_cubiccoex_leiden1896} (see also the review
\cite{crit_sengersexp_physa1976}). It is also well established
fact for a wide variety of the Lennard-Jones fluids with the
short ranged interactions. The results of computer simulations
are well described by the Guggenheim-like expression
\ref{cubicextrapol} (see \cite{pcs_guggenheim_jcp1945}).

The transformation \eqref{projtransfr_nx} states that the shape
of the CC for the molecular fluids is determined by that for
the LG \cite{eos_zenomeglobal_jcp2010}. As is known from the
computer simulations \cite{crit_cross_ising_binder_pre1998} the
crossover to the classical behavior which is characterized by
the parabolic shape of the binodal is absent for the LG with
the nearest neighbors interaction. On the basis of the global
character of the transformation \eqref{projtransfr_nx} one can
assume that the same is true for the molecular fluids with the
short-ranged potentials. In other words, the binodal of the
molecular fluid can be described approximately by the global
cubic dependence similar to \eqref{cubicextrapol} in a broad
temperature interval. In \cite{crit_globalisome_jcp2010} this
statement was demonstrated for 2D and 3D Lennard-Jones fluids.
In particular the binodal of the 2D Lennard-Jones fluid was
obtained as the image of the binodal of the 2D Ising model
given by the Onsager exact solution. In such a case very flat
shape of the binodal dome of 2D fluid is due to the exponent
$\beta =1/8$ and the global nature of the power-like dependence
of the binodal of the 2D LG:
\[x = 1/2\pm f(t)^{1/8}\,,\quad f(t) = 1-\f{1}{{\rm Sinh}(2J/t)}\,.\]
Note that $f(t)$ is the analytic function of the LG temperature
variable $t$.

The classical Landau theory of the phase transitions gives the
general picture which is characterized by the classical
exponents for the critical asymptotics of the thermodynamic
quantities \cite{book_ll5_en}. In particular the dome of the
binodal is given by the the quadratic curve. But it should be
stressed that the Landau theory breaks the continual character
of change of the thermodynamic state in passing through the CP.
Indeed, there is the discontinuity of the specific heat $C_V$
and therefore there is no unique critical state but still there
are two coexisting phases.
\begin{figure}[hbt!]
\center
\subfigure[\,\,LG]{
  \includegraphics[scale=0.75]{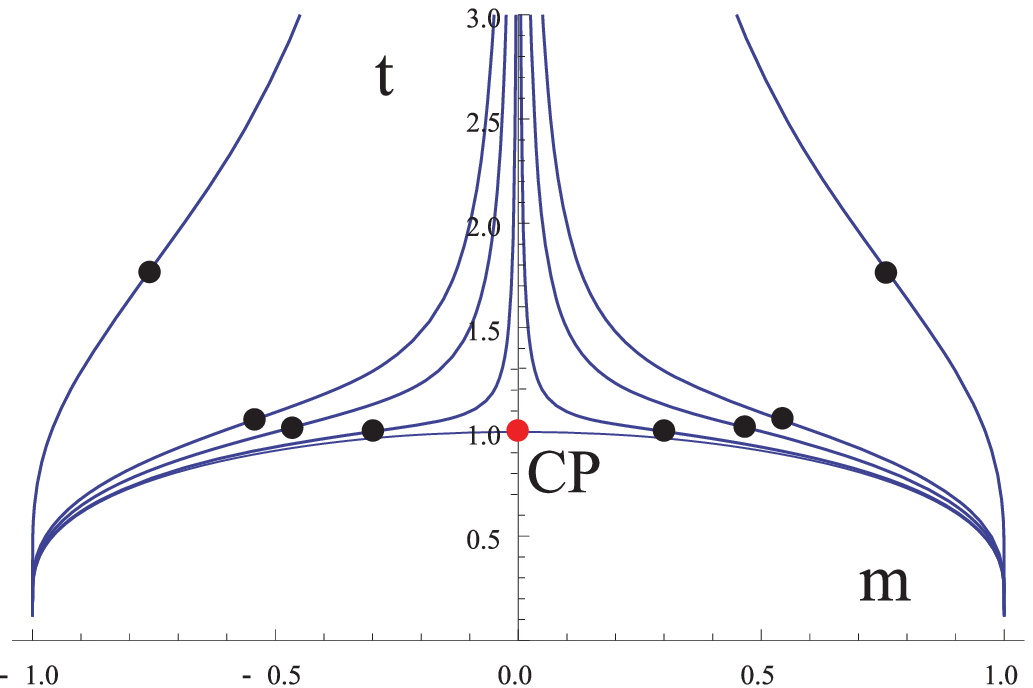}}
\subfigure[\,\,liquid]{
  \includegraphics[scale=0.5]{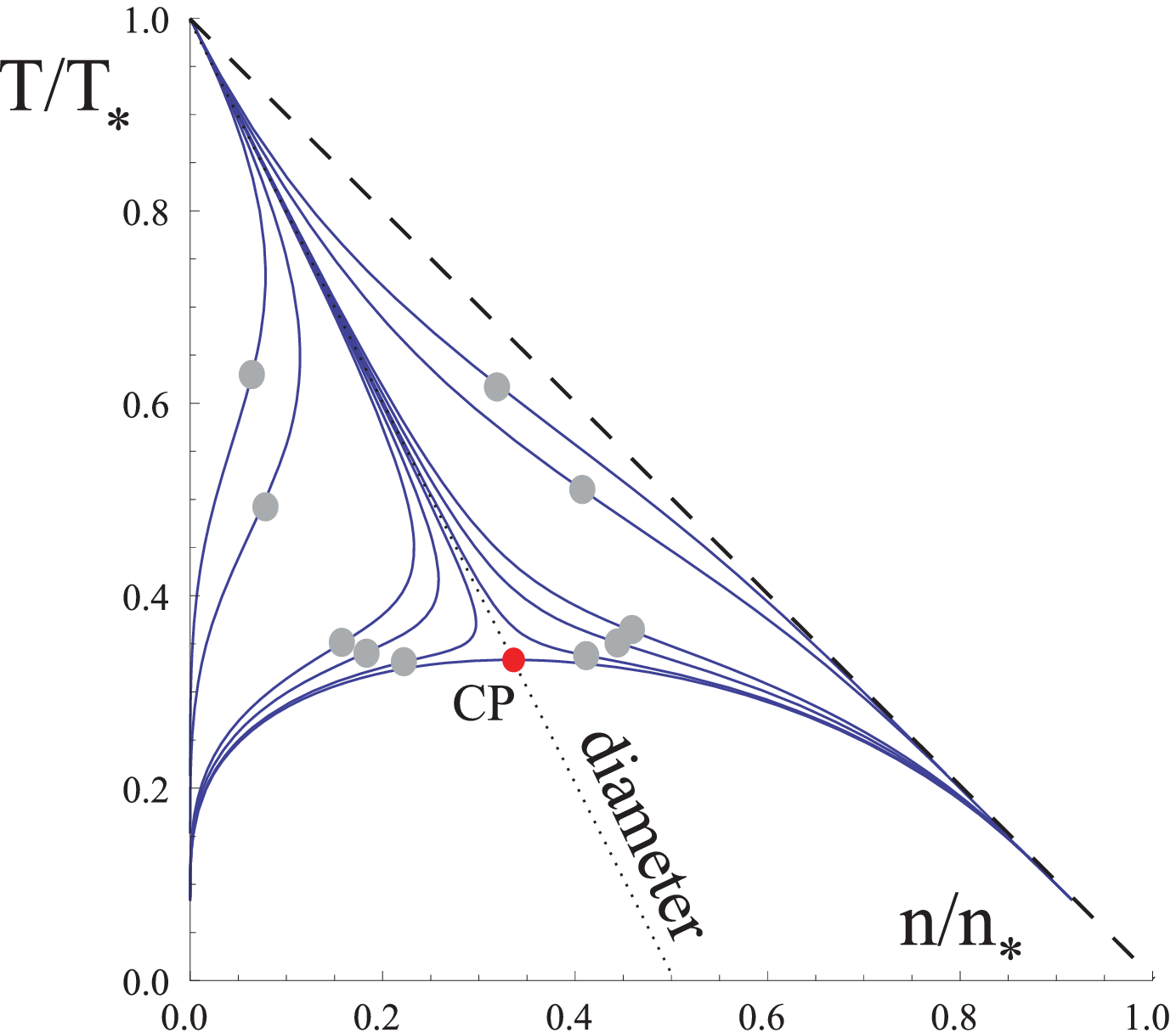}}
  \caption{The phase diagram of the LG (the Ising model)
  and its mapping to $(n,T)$ plane of liquid states with the help of \eqref{projtransfr_nx}.
  The points mark the positions
  of the inflection points (see the text). The red point is the CP. Two families with $h>0$ and $h<0$ of iso-h lines $t(h,m)$ (a) and their images in $n-T$ plane (b) are shown. The binodal is the limiting curve of these families.}\label{fig_ising_1}
\end{figure}
This discontinuity is connected with the assumption of the
existence of the only strongly fluctuating quantity. For the
liquid-vapor critical point such quantity is the density.
Indeed, if the entropy is used as the order parameter then the
finite jump of the specific heat allows to distinguish between
the ordered and the disordered phase at the CP. Besides, the
parabolic shape of the binodal is the consequence of the usage
of analytic EoS with applying the Maxwell construction for the
pair of conjugated variables (e.g. the pressure and the volume)
in the subcritical region. This puts the constraint of
analyticity of the binodal in terms of other pair of the
variables, e.g. the temperature and the entropy. Note that in
the vicinity of the  critical point $t(\tilde{x})$ should be
the even function of the order parameter (see
\eqref{tbx_even}). One can expect that weak divergence of the
specific heat admits the analyticity of the function
$t(\tilde{x})$. Then $t(\tilde{x}) = 1 -
b_0\,\tilde{x}^{2n}+\ldots\,,b_0>0$, where $n>0$ is the
integer. Obviously, this occurs in $d=2$ and $d\ge 4$, where
the specific heat has weak logarithmic divergence or finite
jump with $n=4$ and $n=1$ correspondingly. For $d=3$ the
specific heat diverges more strongly. Therefore one could
expect that the analyticity of $t(\tilde{x})$ breaks in this
case.

\begin{figure}
\center
  \includegraphics[scale=1]{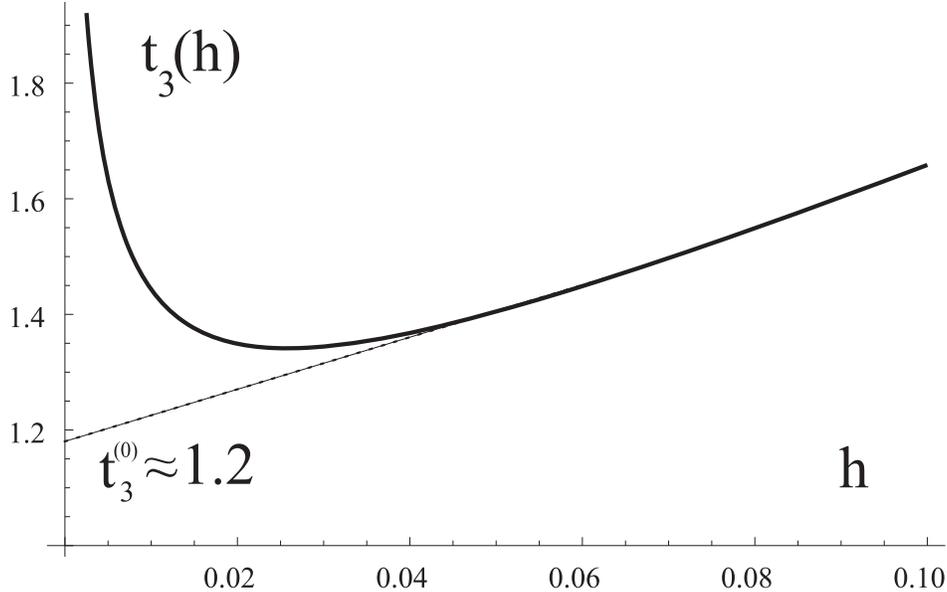}\\
  \caption{The behavior of the coefficient $b(h)$ in the Curie-Weiss approximation. The extrapolation is shown by the dashed line.}\label{fig_t3ising_widomline}
\end{figure}

To clarify where the possibility for the global cubic shape of
the binodal appears let us consider the $t-m$ phase diagram of
the LG (Ising model). We use the Ising-like order parameter $m
= 2x-1$ for the convenience and consider the family of the
\textit{iso}-$h$ curves $t(h,m)$ for the LG. Notably, all these
curves has the inflection points $m_0(h)$.
Fig.~\ref{fig_ising_1} shows the situation for the Curie-Weiss
EoS \eqref{cw_eos}. Obviously, in the vicinity of the
inflection point $m_0(h)$ determined by the condition:
\[
\left.\frac{\partial^2\, t(h,m)}{\partial\, m^2}\right|_{m=m_0}  = 0\,,\]
the function $t(h,m)$ has the cubic form (see
\ref{fig_ising_1}):
\begin{equation}\label{th3}
  t(h,m) = t_0(h) + t_1(h)\,\left(\,m-m_0(h) \,\right) +
  \f{t_3(h)}{6}\,\left(\,m-m_0(h) \,\right)^3+\ldots\,\,,
\end{equation}
where the coefficients $t_1(h)$ and $t_3(h)$ are the odd
functions of $h$. Naturally, the binodal as the line of the
phase equilibrium is the curve which correspond to $h=0$. From
this point of view the binodal consists of two parts. They are
the limiting curves of the families of cubics \eqref{th3} with
$h>0$ and $h<0$ correspondingly (see Fig.~\ref{fig_ising_1}).
Therefore the binodal as the limiting curve for these families
should have the cubic form:
\begin{equation}\label{t3}
  t(0\pm 0,m) = 1 - \f{t^{(0)}_{3}}{6}\,m^2\,|m|+\ldots\,\,.
\end{equation}
or, equivalently
\begin{equation}\label{cubic_binodal}
  m = B_0|\tau|^{1/3} + \ldots\,,
\end{equation}
where
\begin{equation}\label{b0}
  B_0 = \left(\,\f{6}{t^{(0)}_{3}} \,\right)^{1/3}\,.
\end{equation}
In the mean-field approximation $t_{3}(h)\to \infty$ in the CP.
But the analysis of the simplest Curie-Weiss approximation
shows that it is reasonable to use the extrapolated value
$t^{(0)}_{3}$ (see Fig.~\ref{fig_t3ising_widomline}).
From \eqref{b0} we obtain the value of the amplitude for the
binodal opening $B_0\approx 1.7$. Surprisingly, this coincides
with the result of the corresponding value obtained in the
computer simulations for the 3D Ising model
\cite{crit_3disingmc_jmathphys1996}. Under the transformation
\ref{projtransfr_nx} the iso-$h$ lines map onto the
corresponding ones in $n-T$ plane (see Fig.~\ref{fig_ising_1}).
Obviously, the local cubic character similar to \eqref{th3}
near the inflection points takes place. Approaching the CP we
obtain the relation between the amplitudes of the LG and the
continuum fluid:
\begin{equation}\label{blg_bfluid}
  B^{(f)}_0 = (1+z)^{1/3}\,B_0\,.
\end{equation}
In accordance with the results \cite{eos_zenomeglobal_jcp2010}
for the Lennard-Jones fluids, where the attractive part of the
interaction has the ``$-1/r^6$`` behavior, $z=1/2$ in $d=3$.
Then, from \eqref{b0} and \eqref{blg_bfluid} we get:
\begin{equation}\label{b0_fluid}
B^{(LJ)}_0 \approx 1.95\,,
\end{equation}
which is good agreement with the value obtained in computer
simulations $B_0\approx 1.92 \div 2.0$ for potentials of
``$6-n$`` type \cite{crit_liqvamiepotent_jcp2000,
*crit_longrangecampatey_jcp2001,*eos_ljfluid_jcp2005}.

This allows to treat the value $\beta = 1/3$ as the
corresponding exponent which conforms with the continuity of
the critical state for $d=3$ for the systems isomorphic to the
LG. Other critical indices, except the small ones $\alpha$ and
$\eta$ which in the mean-field approximation are zeroth, can be
obtained via the standard thermodynamic stability and scaling
reasonings, so that $\nu = 2/3,\,\gamma =4/3,\,\delta =5$ etc.

\section{Conclusions}
In this paper the explicit relations between the basic
thermodynamic functions of the LG and the continuum fluid is
derived within the global isomorphism approach proposed in
\cite{eos_zenome_jphyschemb2010}. These relations allow to
obtain the information about liquid state directly from the EoS
for the lattice model. It is quite remarkable since the lattice
models do not contain the translational degrees of freedom.
Nevertheless, as it follows from the results
\cite{crit_globalisome_jcp2010,eos_zeno_lattice2real_jpcb2010}
this mapping gives rather good description for the LJ fluids.
In particular the stated relations can be used to connect the
computer simulations for the lattice models with those for the
continuum fluids (see ~\cite{crit_globalisome_jcp2010}). Within
such an approach the splitting between critical isochore and
the diameter, the Zeno-line and the tangent to the binodal at
$T\to 0$ is naturally described. The conservation of linear
character of the diameter allows to obtain simple form of the
transformation between the phase diagram of the LG and the
continuum fluid. Obviously, the value of the critical density
is the most sensitive to such approximation. But for the
computer simulations the deviations from the law of rectilinear
diameters are small to observe them near the CP
\cite{book_frenkelsimul}. This explains rather good agreement
of the estimates for the locus of the CP based on the
parametrization \eqref{projtransfr_nx} for the Lennard-Jones
fluids with the results of the computer simulations
\cite{eos_zenogenpcs_jcp2010}.

We believe that the approach proposed could be useful in
studies of the supercritical behavior basing on the EoS of the
lattice models especially for water
\cite{fwidomline_jpcm2007,water_widomline_jcp2010} where a lot
of lattice models, including the models with H-bond are known.
In view of the results of work \cite{crit_widomline_nature2010}
it would be interesting to generalize the proposed approach in
order to search the correspondence between the dynamic response
functions of fluids and their lattice analogues. This poses the
question about the choice of the adequate lattice model of the
proper geometry and the type of the interaction which realizes
the isomorphism with the corresponding fluid.

\section*{Acknowledgements}
The authors cordially thank Prof. N. Malomuzh for fruitful
discussion of the obtained results and valuable comments.
%

\end{document}